\documentclass[ reprint, amsmath,amssymb, aps,prx]{revtex4-2}

\usepackage{graphicx}
\usepackage{dcolumn}
\usepackage{bm}
\usepackage{amsmath}
\usepackage{amssymb}
\usepackage[usenames,dvipsnames]{xcolor} 
\usepackage{mathtools}
\usepackage[colorlinks=true,citecolor=blue,linkcolor=blue]{hyperref}
\usepackage[caption=false]{subfig}
\usepackage{bbm}
\usepackage{braket}
\usepackage{comment}
\usepackage{hyperref}
\usepackage{upgreek}
\usepackage{esint}
\usepackage{soul}
\usepackage[normalem]{ ulem } 
\usepackage{floatrow}
\usepackage{esint}
\usepackage{tikz}
\usepackage{wrapfig}

\begin{document}

\title{Internal entropy from heat current}

\author{Noam Schiller$^1$}
\author{Hiromi Ebisu$^{1,2,3}$}
\author{Gil Refael$^4$}
\author{Yuval Oreg$^1$}
\affiliation{
\small{$^1$Department of Condensed Matter Physics, Weizmann Institute of Science, Rehovot 7610001, Israel \\
$^2$Department of Physics and Astronomy, Rutgers University, Piscataway, NJ 08854, USA \\ 
$^3$ Yukawa Institute for Theoretical Physics, Kyoto University, Kyoto 606-8502, Japan\\
$^4$ Department of Physics and Institute for Quantum Information and Matter, California Institute of Technology, Pasadena, CA 91125
}}

\date{\today}

\begin{abstract}
  We demonstrate that the effective internal entropy of quasiparticles within the non-Abelian fractional quantum Hall effect manifests in the heat current through a tunneling barrier. We derive the electric current and heat current resulting from voltage and heat biases of the junction, taking into account the quasiparticles' internal entropy. We find that when the tunneling processes are dominated by quasiparticle tunneling of one type of charge, the effective internal entropy can be inferred from the measurement of the heat current and the charge current. Our methods may be used to conclusively identify non-Abelian quasiparticles, such as the anyons that emerge in the $\nu = 5/2$ fractional quantum Hall state.
\end{abstract}

\maketitle

\textit{Introduction.}--- One of the most fundamental striking features of matter's topological phases is how entropy uniquely characterizes different states. For example, non-Abelian topological states, such as the $\nu=5/2$ fractional quantum Hall (FQH) state \cite{moore_nonabelions_1991,levin_particle-hole_2007,lee_particle-hole_2007,stern_anyons_2008,nayak_non-abelian_2008}, are portrayed by degenerate ground state manifolds, or large Hilbert spaces.

In particular, a system of $N$ well-spaced anyons of type~$\alpha$ is described by a Hilbert space of asymptotic dimension $\sim d_\alpha^{N}$ at the limit of large $N$ \cite{nayak_non-abelian_2008}, where $d_\alpha$ is defined as the quantum dimension of the anyon. This allows a definition of an effective ``internal entropy per particle", given immediately by $s_\alpha \equiv S/N = \log (d_\alpha)$. Remarkably, the same quantum dimension governs the topological entanglement entropy, a size-independent entanglement entropy deficit of a segment of a topological phase. The topological entanglement entropy is given by~\cite{Levin_detecting_2006,kitaev_topological_2006} $-\gamma = -\log \sqrt{\sum_{\alpha}d_{\alpha}^{2}}$, where the summation is over all anyons which are supported by the system.

However, measuring the internal entropy of non-Abelian quasiparticle excitations is highly challenging as their contribution to extensive system properties (such as, for instance, heat capacity) is minute. There have been proposals to measure the internal entropy of non-Abelian anyons at the $\nu=5/2$ systems by measuring thermopower~\cite{yang_thermopower_2009}, the Ettingshausen effect~\cite{hou_ettingshausen_2012}, and by measuring the bulk magnetization density~\cite{cooper_observable_2009} and relating it to entropy through Maxwell relations. 

The ability to accurately measure the charge of a quantum dot enables the successful use of the Maxwell relation to measure the entropy in a quantum dot due to the spin degree of freedom~\cite{hartman_direct_2019}. Similar measurements were proposed as methods to be used to observe the entropy of Majorana zero modes or the topological entanglement entropy of further topological states \cite{sela_detecting_2019,sankar_measuring_2022}.

 \begin{figure}
     \centering
\tikzset{every picture/.style={line width=0.75pt}} 
\begin{tikzpicture}[x=0.75pt,y=0.75pt,yscale=-1,xscale=1]
\draw  [color={rgb, 255:red, 255; green, 255; blue, 255 }  ,draw opacity=1 ][fill={rgb, 255:red, 173; green, 216; blue, 230 }  ,fill opacity=0.15 ] (100,110) -- (400,110) -- (400,200) -- (100,200) -- cycle ;
\draw [fill={rgb, 255:red, 255; green, 255; blue, 255 } ,fill opacity=1 ][line width=2.25]    (100,110) .. controls (110,110) and (190,111) .. (200,111) .. controls (250,110) and (240,141) .. (250,140) .. controls (260,140) and (250,110) .. (300,110) .. controls (310,110) and (390,110) .. (400,110) ;
\draw [shift={(360,110)}, rotate = 180] [fill={rgb, 255:red, 0; green, 0; blue, 0 }  ][line width=0.08]  [draw opacity=0] (15,-4) -- (0,0) -- (15,4) -- cycle    ;
\draw [shift={(150,110.5)}, rotate = 180] [fill={rgb, 255:red, 0; green, 0; blue, 0 }  ][line width=0.08]  [draw opacity=0] (15,-4) -- (0,0) -- (15,4) -- cycle    ;
\draw [fill={rgb, 255:red, 255; green, 255; blue, 255 }  ,fill opacity=1 ][line width=2.25]    (400,200) .. controls (390,200) and (310,200) .. (300,200) .. controls (250,200) and (260,170) .. (250,170) .. controls (240,170) and (250,200) .. (200,200) .. controls (190,200) and (110,200) .. (100,200) ;
\draw [shift={(130,200)}, rotate = 0] [fill={rgb, 255:red, 0; green, 0; blue, 0 }  ][line width=0.08]  [draw opacity=0] (15,-4) -- (0,0) -- (15,4) -- cycle    ;
\draw [shift={(350,200)}, rotate = 0] [fill={rgb, 255:red, 0; green, 0; blue, 0 }  ][line width=0.08]  [draw opacity=0] (15,-4) -- (0,0) -- (15,4) -- cycle    ;
\draw    (236,130) .. controls (220,140) and (220,170) .. (240,180) ;
\draw [shift={(241,180.5)}, rotate = 210] [fill={rgb, 255:red, 0; green, 0; blue, 0 }  ][line width=0.08]  [draw opacity=0] (9,-4.5) -- (0,0) -- (9,4.5) -- cycle    ;
\draw [shift={(236,130)}, rotate = 150] [color={rgb, 255:red, 0; green, 0; blue, 0 }  ][fill={rgb, 255:red, 0; green, 0; blue, 0 }  ][line width=0.75]      (0, 0) circle [x radius= 4, y radius= 4]   ;
\draw (220,155) node   {$p$};
\draw    (264,130) .. controls (280,140) and (280,170) .. (260,180) ;
\draw [shift={(263,179)}, rotate = 150] [fill={rgb, 255:red, 255; green, 255; blue, 255 }  ][line width=0.75]      (0, 0) circle [x radius= 4, y radius= 4]   ;
\draw [shift={(260,128)}, rotate = 30] [fill={rgb, 255:red, 0; green, 0; blue, 0 }  ][line width=0.08]  [draw opacity=0] (9,-4.5) -- (0,0) -- (9,4.5) -- cycle    ;
\draw (270,155) node  [font=\normalsize]  {$h$};
\draw    (272,120) .. controls (280,120) and (295,130) .. (295,151) ;
\draw [shift={(295,151)}, rotate = 90] [fill={rgb, 255:red, 255; green, 255; blue, 255 }  ][line width=0.75]      (0, 0) circle [x radius= 4, y radius= 4]   ;
\draw [shift={(270,117.5)}, rotate = 30] [fill={rgb, 255:red, 0; green, 0; blue, 0 }  ][line width=0.08]  [draw opacity=0] (9,-4.5) -- (0,0) -- (9,4.5) -- cycle    ;
\draw    (272,190) .. controls (280,190) and (295,180) .. (295,159) ;
\draw [shift={(295,159)}] [color={rgb, 255:red, 0; green, 0; blue, 0 }  ][fill={rgb, 255:red, 0; green, 0; blue, 0 }  ][line width=0.75]      (0, 0) circle [x radius= 4, y radius= 4]   ;
\draw [shift={(270,192.5)}, rotate = 330] [fill={rgb, 255:red, 0; green, 0; blue, 0 }  ][line width=0.08]  [draw opacity=0] (9,-4.5) -- (0,0) -- (9,4.5) -- cycle    ;
\draw (307,155) node  {$hp$};
\draw    (228,190) .. controls (210,190) and (200,180) .. (200,158) ;
\draw [shift={(200,155)}, rotate = 90] [fill={rgb, 255:red, 0; green, 0; blue, 0 }  ][line width=0.08]  [draw opacity=0] (9,-4.5) -- (0,0) -- (9,4.5) -- cycle    ;
\draw [shift={(225,190)}] [fill={rgb, 255:red, 255; green, 255; blue, 255 }  ][line width=0.75]      (0, 0) circle [x radius= 4, y radius= 4]   ;
\draw    (228,120) .. controls (210,120) and (200,130) .. (200,152) ;
\draw [shift={(200,155)}, rotate = 269.46] [fill={rgb, 255:red, 0; green, 0; blue, 0 }  ][line width=0.08]  [draw opacity=0] (9,-4.5) -- (0,0) -- (9,4.5) -- cycle    ;
\draw [shift={(225,120)}] [color={rgb, 255:red, 0; green, 0; blue, 0 }  ][fill={rgb, 255:red, 0; green, 0; blue, 0 }  ][line width=0.75]      (0, 0) circle [x radius= 4, y radius= 4]   ;

\draw (190,155) node [font=\normalsize]  {$ph$};
\draw [line width=2]    (250,140)  -- (250,170) ;
\draw [shift={(250,160)}, rotate = 270] [fill={rgb, 255:red, 0; green, 0; blue, 0 }  ][line width=0.08]  [draw opacity=0] (9,-4.5) -- (0,0) -- (9,4.5) -- cycle  ;
\draw (243,160) node   {$\it \mathbf{e^{\!*}}$};
\draw (140,130) node   [font=\Large]  {$\nu,\delta $};
\draw (336,178) node   [font=\Large]  {$ -\mathbf{e^{\!*}} \sigma$};

\draw [shift={(348,170)}] [fill={rgb, 255:red, 255; green, 255; blue, 255 }  ][line width=0.75]      (0, 0) circle [x radius= 4, y radius= 4];
\draw (360,140) node   [font=\Large]  {$\mathbf{e^{\!*}} \sigma$};
\draw [shift={(360,150)}] [fill={rgb, 255:red, 0; green, 0; blue, 0 }  ][line width=0.75]      (0, 0) circle [x radius= 4, y radius= 4];

\draw   (100,110) -- (120,110) -- (120,130) -- (100,130) -- cycle ;

\draw (100,112) node [anchor=north west][inner sep=0.75pt]  [font=\large,rotate=0]  {$\text{S}_{\text{R}}$};

\draw   (380,110) -- (400,110) -- (400,130) -- (380,130) -- cycle ;

\draw (378,112) node [anchor=north west][inner sep=0.75pt]  [font=\large,rotate=0]  {$\text{D}_{\text{R}}$};

\draw   (100,180) -- (120,180) -- (120,200) -- (100,200) -- cycle ;

\draw (98,182) node [anchor=north west][inner sep=0.75pt]  [font=\large,rotate=0]  {$\text{D}_{\text{L}}$};

\draw   (380,180) -- (400,180) -- (400,200) -- (380,200) -- cycle ;

\draw (381,182) node [anchor=north west][inner sep=0.75pt]  [font=\large,rotate=0]  {$\text{S}_{\text{L}}$};
\end{tikzpicture}
\caption{Four possible tunneling processes between the top and bottom edges of a non-Abelian state (shaded area). All processes result in transferring an anyon charge $e^*=e/4$ from top to bottom (straight arrow), however, the entropy transfer is different in each process. For example, in the non-Abelian $\nu=5/2$ state, the $p$ process transfers a quasiparticle with charge $e^*$ together with its internal entropy $s_\sigma =\log(\sqrt{2})$ from top to bottom. The $ph$ process has an identical charge transfer, but it takes entropy from both the top and the bottom edges and heats the bulk. The $hp$ and the $h$ processes are similar and are discussed in the text. \label{fig:sys}\vspace{-.7cm}}
 \end{figure}
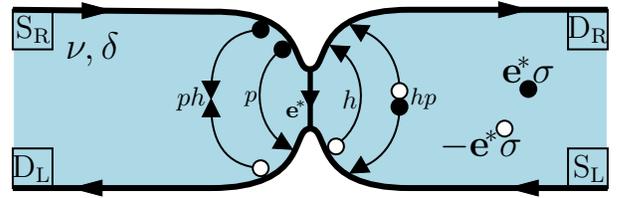
 
The current work proposes to deduce non-Abelian anyons' entropy through heat and charge transport measurements. We consider a system at the fractional quantum hall regime hosting a non-Abelian state at, for example, filling~$\nu=5/2$. The system's bulk is bounded by a top and a bottom edges with chemical potentials $\mu_{t/b}$ and temperatures $T_{t/b}$, and allow rare tunneling events of anyons between them, see Fig.~\ref{fig:sys}.

\begin{figure}

\input{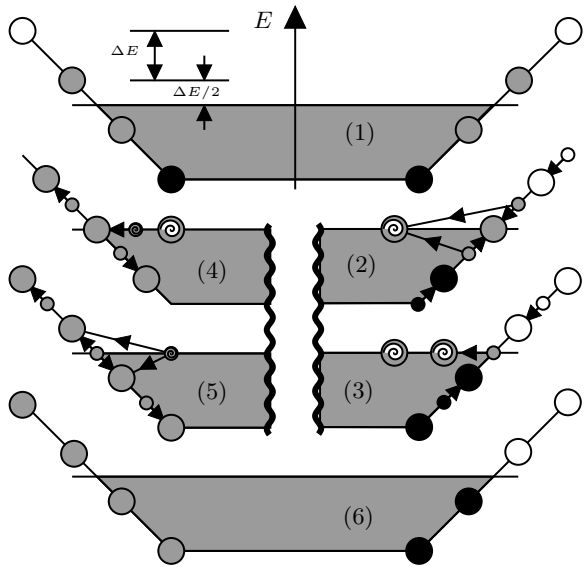}

\caption{\label{fig:spectralflow} Transfer of quasiparticles (vortices) from the right movers' edge to the left movers' edge leads to a spectral flow, heat current, and entropy transfer.  The electric charge carried by the vortices is irrelevant to the following discussion. \textbf{(1)}~Due to the finite temperature, the BdG states near the Fermi level are partially filled (grey shade), and those far away are either empty (white, above the Fermi level) or full (black, below the Fermi level). In the example we consider in the figure, the initial state does not have Majorana zero modes; one should average over all possible states. \textbf{(2)} When a vortex enters the Hall constriction bulk (from the vacuum) through the right movers' edge, its $\sigma$ twist operator changes the boundary conditions and produces a spectral flow of the energies of the BdG chiral edge states towards~$E=0$ by $\frac{1}{2}\Delta E=\frac{1}{2}\frac{2\pi v}{L}$. A Majorana zero-mode is formed on the edge and in the vortex. \textbf{(3)} When two vortices enter the constriction (through the right movers' edge), they carry with them a full complex fermion BdG state, leaving the edge spectrum unchanged but causing all particle-BdG energies to flow down by $\Delta E= \frac{2\pi v}{L}$ and the hole-BdG states energies up by $\Delta E$.  Since the empty particle states flow towards the Fermi level from the top and the full hole states from the bottom, the edge cools down. Because the temperature is larger than the (equal to zero) level spacing at finite temperature, the fermion state of the two fermions is full or empty with equal probabilities, so an entropy of $\log (2)$ was taken away from the right movers' edge. \textbf{(4)} An opposite process occurs near the right movers' edge. When a vortex leaves the bulk through the left movers' edge, it pushes the BdG states away from the Fermi level, bringing in a  Majorana zero mode to it. \textbf{(5)} Two vortices that leave the constriction deposit a complex fermion state, which could be either full or empty, on the left movers' edge. \textbf{(6)} Heat was transferred from the right movers' to the left movers' edge. The entropy transfer was  $s_0=\log \sqrt{2}$ per vortex.\label{vortex-passage} }
\end{figure}

The non-Abelian $\nu=5/2$ state supports bulk anyons with positive and negative charges $\pm |e^*|$ and quantum dimension~$s_0=\log(\sqrt2)$. As these anyons pass from top to bottom, they also alter the entropy of the system's edges. Based on the spectral flow analysis detailed below (see Fig.~\ref{fig:spectralflow}), we find several processes that transfer the same charge between the edges, yet change the edge entropy in different ways. These processes are illustrated in Fig.~\ref{fig:sys}. (i)~In the $p$ process, an anyon with positive charge tunnels from top to bottom, transferring a charge $|e^*|$ and entropy $s_0$ from top to bottom. (ii)~The $h$ process transfers negative charge and entropy from bottom to top. (iii)~In the $hp$ process, positive charge is transferred from the bulk to the bottom edge and negative charge from the bulk to the top edge. Entropy is transferred from the bulk to the top and bottom edges, heating both. (iv)~In the $ph$ process, positive charge is transferred from the top edge to the bulk and negative charge from the bottom to the bulk, where they annihilate each other. Entropy flows from the edges to the bulk.

Below, we evaluate the contribution of each process to the top-to-bottom electric current $I$ and to the heat current, $I_Q$, arriving to the bottom edge. The heat current consists of the standard Joule heating energy current $I_{\rm J}$, as well as a contribution due to the internal entropy of the quasiparticles $I_{\rm S}$. We find that, in the shot noise limit, when $V \gg T_R,T_L$, and when one type of quasiparticle dominates charge transfer across the constriction, there is a universal relation between $I$ and $I_{\rm Q}$:
\begin{equation}
    \label{eq:IQ}
    I_{\rm Q}=I_{\rm J}+I_{\rm S}=\frac{1}{2} I V+ \frac{I}{e^{*}} T_L s_0 + \mathcal{O} \left( \frac{T_R^2}{V},\frac{T_L^2}{V}\right).
\end{equation}
The Joule heating term $I_{\rm J}=\frac{1}{2} I V$ has a $1/2$ factor similar to the one in electron transport, arising from an underlying Fermi sea. The second term, $I_{\rm S} =I_N T_L s_0$, where $I_N$ is particle current, arises due to the internal entropy of the quasiparticles carried in the tunneling process. The heat current due to the internal entropy is simply the number of quasiparticles moving from the right movers' edge to the left movers' edge, multiplied by the internal quasiparticle entropy $s_0$ times the temperature at the left-movers edge. If there is only one type of quasiparticle arriving at the left movers' edge then $I_N=I/e^*$.

Eq.~(\ref{eq:IQ}) describes a relationship between the heat and charge currents. It should be possible to infer these currents through the measurement of the heat and charge currents in different locations of the constriction, employing charge and energy conservation.

Determining whether the $p$ or the $h$ processes are dominant in a specific experiment is challenging, as it depends on many details, such as material properties and gate configuration. We may speculate, for example, that when the bulk and lower edge are grounded, and the upper (right-moving) edge is connected to a source reservoir at the left upper corner (${\rm S_L}$ in Fig.~\ref{fig:sys}) with an elevated potential, the transfer of charge from top to bottom likely occurs through virtual processes that involve an additional positive charge in the bulk. In this scenario, $p$ processes are more dominant than $h$ processes, and Eq.~(\ref{eq:IQ}) is valid. If, on the other hand, the top edge and the system bulk are grounded, and a reservoir is held at a negative potential at the bottom-right corner, we expect that $h$ processes will dominate the transfer of charge and heat. As a result, negative charge current and positive heat current will flow from top to bottom, consistent with Eq.~(\ref{eq:IQ}). This results in a positive electric current from top to bottom, as in the particle-dominant case, but a heat current that is flowing in the opposite direction. 

We note that some quantum Hall states have neutral modes that propagate in the opposite direction of the charge edge modes, known as upstream neutral modes. When a particle tunnels from top to bottom, it may lose some of its energy to the neutral modes, potentially affecting the transfer of heat from top to bottom. One, however, can estimate the heat carried by the tunneling particles by measuring the heat loss on the top edge due to the tunneling event at the point-contact. Since the majority of the heat is carried by the downstream mode at the upper edge, the impact of neutral modes is expected to be minimal.

Below, we discuss how entropy is transferred with the tunneling of quasi-particles using a spectral flow argument. Our need to discuss these processes using that argumentation reveals that the non-Abelian conformal field theory~(CFT) of Ising anyons edges is incomplete. Since the CFT does not retain short time information about dynamical processes it is insensitive to the direction in which spectral changes happen. To wit, the same twist operators are used to account for the addition of a positive charge and the reduction of a negative charge. As a result, it cannot distinguish between an increase or decrease of entropy on the edge due to the internal entropy of the particles. To address this issue, we add `by hand' the difference between the two processes. We leave a technical discussion of this approach to future work.

\textit{Entropy and heat transfer in the $\nu=5/2$ states due to the $\sigma$-particles.}---Our main goal is to clarify how the anomalous quantum dimension of the quasiparticle excitations of the $\nu=5/2$ FQH state play a role in the heat transfer accross a constriction, and how it could be measured. To make progress, we concentrate on the transport events expected to dominate heat and charge transport across a constriction. We do so from a perspective of a p-wave superconductor \cite{fendley_boundary_2009} 
which supports vortices with Majorana zero modes~\footnote{In contrast to the p-wave superconductor, which has classical vortices, the non-Abelian quantum hall states have quantum vortex-like quasiparticles carrying zero modes}.

The dominant process in a point contact between two $\nu=5/2$ edges is the tunneling of an $\pm e/4$ charge quasiparticle.  For our purposes, the important point of emphasis is that each $\pm e/4$ quasiparticle contains a twist field, denoted by a $\sigma$ operator, which changes the boundary condition on the Majorana fermion mode $\psi$ propagating along the edge. Each $\sigma$ crossing an edge changes the boundary conditions of the edge it crosses by a $(-1)$ multiplicative factor. This composite object is nothing more than a vortex crossing into the p-wave paired Pfaffian state. The $\sigma$'s carry with them additional entropy by virtue of their fusion rule: 
\begin{equation}
    \sigma\times\sigma=1+\psi
\end{equation}
Namely, two $\sigma$'s give rise to a single fermionic state, which could be either empty (1) or occupied ($\psi$). Leaning back on the vortex picture, we can easily interpret this fusion rule. As in a p-wave superconductor, each vortex contains a Majorana-like zero energy BdG state at its core, and pairs of vortices together give rise to a degenerate fermionic state that could be either full or empty. As pairs of vortices pass from one side of the constriction to the other, they must pass entropy associated with the degenerate fermionic state with them. 

The entropy transfer in vortex-passage events is connected to changes in the edges of the Hall constriction. The edges have a charge mode, from which the charge of the vortex is taken, and more importantly, a chiral Majorana fermion mode (not to be confused with the Majorana-like BdG zero energy states at the core of p-wave vortices), whose field we denote as $\chi$.
When the boundary conditions of the $\psi$ field are periodic, the $\psi$ eigenmodes have energies $E=2\pi n v/L$ with $n\in {\cal Z}$ where $v$ is the edge velocity, and $L$ the edge length. Note that these are Bogoliubov modes of the $\psi$ field. They imply single particle states with energies $E=2\pi |n| v/L$. Particularly, this includes $E=0$. If the boundary conditions change to anti-periodic, the BdG edge energy (single-particle) spectrum changes to $E=2\pi v (|n|+1/2)/L$, removing the zero-energy Majorana state.

Relying on spectral flow and conservation of spectral weights, we can now follow the process of vortices entering and leaving the $\nu=5/2$ constriction (see Fig. \ref{vortex-passage}). As a vortex, along with its $\sigma$ operator, enters the constriction (from the vacuum), it changes the boundary conditions of the edge and thus changes the energies of the BdG edge states. Since we know that the vortex carries with it a Majorana zero-energy state, we conclude that the state was peeled off the edge, and, therefore, the edge spectrum was sucked in towards zero with each particle (hole) state dropping (rising) in energy by 
\begin{equation}
   \frac{1}{2} \Delta E=\frac{1}{2} \frac{2\pi v}{L}.
\end{equation}
The next vortex to enter will inevitably repeat this process: draw the BdG edge spectrum towards zero by $\frac{1}{2}\Delta E=\frac{1}{2} \frac{2\pi v}{L}$, leaving behind a spectrum that again does not include a Majorana zero energy state.  The first and second vortices carry between them a complex fermionic state that could have been empty or full, indicating an entropy of $\log 2$ that they draw from the edge.

The same process, but in reverse, happens as the vortices reach the other side of the constriction. The first vortex brings a Majorana zero-mode with it which pushes the edge spectrum up by $\frac{1}{2}\Delta E= \frac{1}{2}\frac{2\pi v}{L}$, when the second vortex arrives it partners with the zero-mode on edge to produce the new $E=\frac{1}{2}\frac{2\pi v}{L}$ BdG complex fermionic state. 

Crucially, as vortices flow, every pair of vortices arriving at the second edge brings with them a complete complex fermionic state, which could be either empty or occupied. They push the spectrum of the receiving edge away from zero by $\Delta E=\frac{2\pi v}{L}$, and introducing a pair of particle and hole states, adding an entropy of $\Delta S_{2\sigma}= \log 2$ for the 2 $\sigma$'s. In this picture, the passage is considered adiabatic  (namely, the virtual tunneling process is slower in imaginary time than the inverse of the gap of the Hall state) so that the occupation, and therefore entropy of the rest of the BdG edge states on the receiving end, remain the same. 
The change in heat in the receiving (left-moving) end is therefore: 
\begin{equation}
    \Delta Q_{\sigma}=\frac{1}{2} T_L \log 2.
\end{equation}
It is this heat that we hope the measurements suggested in this manuscript will reveal. 

 Before we close, we should add a short note on the parity of each edge. We point out that the $\psi$ states, which determine the parity of the edges, are encoded in the occupations of the low-lying BdG states of each edge. Indeed, if a pair of vortices transfers a state $\psi$ from the top edge to the bottom edge, the parity of both edges changes; namely, the parity of the set of low-lying BdG states in each edge is changed. As vortices continue to move between the edges, low-energy BdG states move closer to the $E=0$ line and have an opportunity to transfer their parity to the other edge. 

The excess entropy that is described by the spectral flow arguments above does not require adiabaticity. Non-equilibrium aspects of the tunneling process are fully captured in the low-energy theory of each edge. The entropy associated with the non-Abelian nature of the tunneling objects is captured by the boundary condition change and the subsequent spectral flow regardless of its speed.

\textit{Quantum rate equations for the heat and the charge current at $\nu=5/2$.}---
In this section, we present a formal calculation of the heat and charge currents at $\nu=5/2$.

The edge Hamiltonian that captures all proposed underlying orders which may describe the $\nu=5/2$ state includes two integer edge modes, corresponding to the fully occupied Landau levels, and some combination of charged boson mode and neutral Majorana modes. We henceforth denote the boson mode on the right-/left-moving edge as $\phi_{R/L}$. These correspond to charge density on the edges via the relation $\rho_{R/L}=\frac{e}{2\pi}\partial_x \phi_{R/L}$, satisfying the commutation relations $\left[\phi_{R/L}(x),\phi_{R/L}(x^\prime)\right] = \pm i\frac{\pi}{2} \textrm{sgn}\left(x-x^\prime\right)$. We denote the neutral Majorana modes as $\chi_{R/L,j}$. These modes are self-adjoint and satisfy fermionic anticommutation relations, $\left \{ \chi_{R/L,j}(x),\chi_{R/L,k}(x^\prime) \right\} = 2 \delta_{j,k} \delta \left(x-x^\prime \right)$.

Neglecting the two full Landau levels, the single-edge Hamiltonian (with $r=R/L$) is then given by
\begin{align}
    \label{eq:Hamiltonian_five_halves}
    H_r & =  \int dx \left( \frac{v_{rc}}{2\pi}\left( \partial_x \phi_r \right)^2 +
    \sum_{j=1}^{N_\chi} i \chi_{rj} v_{rn,j} \partial_x \chi_{rj} \right).
\end{align}
where $v_{rc}$ and $v_{rn,j}$ are the edge velocities of the charged and the $j$th neutral mode, respectively. Different topological orders are characterized by the number $N_\chi$ \footnote{In particular, the Pfaffian state has a single Majorana fermion with a positive velocity; the anti-Pfaffian has three Majorana fermions with negative velocities; and the PH-Pfaffian state, which thermal conductance measurements have so far supported \cite{banerjee_observed_2017,banerjee_observation_2018,dutta_distinguishing_2022,dutta_isolated_2022} despite scant numerical evidence, has one Majorana fermion with a negative velocity.} and propagation direction of neutral modes \cite{di_francesco_conformal_2005,nayak_non-abelian_2008,fendley_edge_2007,yang_influence_2013}. 

These edge theories support edge excitations which correspond to primary fields of the underlying 
CFT~\cite{ginsparg_applied_1988,di_francesco_conformal_2005,fendley_edge_2007}. We focus here only on the most relevant quasiparticle type, corresponding to the smallest scaling dimension $\delta$, which should dominate tunneling through the QPC.
The tunneling Hamiltonian at $x=0$  between the edges is
\begin{equation}
    \label{eq:tunneling_hamiltonian}
    H_T = \Gamma \Psi^\dagger_R(x=0) \Psi_L(x=0) + \textrm{h.c.},
\end{equation}
where $\Gamma$ is a tunneling amplitude, assumed to be real.
The annihilation and creation operators of this quasiparticle are given by
\begin{align}
    \label{eq:excitations}
    \Psi_{R/L} \sim \sigma_{R/L} e^{\frac{i}{2} \phi_{R/L}};
    \quad
    \Psi_{R/L}^\dagger \sim \sigma_{R/L} e^{-\frac{i}{2} \phi_{R/L}}.
\end{align}

The exponential of the boson field indicates that this quasiparticle should have a charge of $e^* = e/4$. This prediction has been corroborated in shot noise experiments~\cite{dolev_observation_2008}. More importantly, $\sigma$ is the twist field, which changes the boundary condition of all neutral Majorana fermions when it crosses the edge \cite{ginsparg_applied_1988,di_francesco_conformal_2005,nayak_non-abelian_2008}. The non-Abelian nature of this quasiparticle is best observed in the fusion rules of the underlying CFT,
$\sigma \times \sigma = I + \psi$, which shows that two twist fields fuse into either the identity channel or a fermionic channel. The internal entropy of the quasiparticle $\Psi$ is then $\log \sqrt{2}$, as it consists of a single $\sigma$ field. 

The form of Eq.~\eqref{eq:excitations} shows us an oddity of these quasiparticles, which is that both $e^*=e/4$ quasiparticles and $e^*=-e/4$ quasiholes consist of the same non-Abelian portion, in a single $\sigma$ field. The two terms in Eq.~\eqref{eq:excitations} could describe equally well the tunneling of a quasihole, or of a quasiparticle. 
Also, tunneling of either a quasiparticle or a quasihole carry an internal entropy of $\log \sqrt{2}$, and raise the heat in the reservoir upon arrival and equilibration. Thus, as discussed in detail in the previous section,  we expect the creation of a quasiparticle and the annihilation of a quasihole to be \textit{experimentally distinct} in entropy measurements. This distinction is absent from the operators of Eq.~\eqref{eq:excitations}, or the tunneling Hamiltonian Eq.~\eqref{eq:tunneling_hamiltonian}.

In App.~\ref{app:quantum}, we show, using a straightforward perturbation theory in the tunneling $\Gamma$, that the charge current~$I$ and the energy current $I_J$, due to events with charge transfer $e^*$ are given by:
\begin{subequations}
    \label{eq:generalcurrents}
    \begin{align}
        I & = 2 i e^* \Gamma^2 \int_{-\infty}^\infty d\tau \sin \left(e^* V \tau \right) G_R(\tau) G_L(\tau), \\
        I_{\rm J} & = 2 i \Gamma^2 \int_{-\infty}^\infty d\tau \cos \left(e^* V \tau \right) G_R(\tau) \partial_\tau G_L(\tau),
    \end{align}
\end{subequations}
where $G_r(\tau)$ is a single-particle Green's function. We use the operators of Eq.~\eqref{eq:excitations} to calculate the correlation functions that appear in Eq.~\eqref{eq:generalcurrents}. We further show that, when tunneling is dominated by a single quasiparticle type, $I_S = \left(T_L s_0/e^*\right) I$.
Using known values for correlations of the boson fields~\cite{giamarchi_quantum_2003,wen_quantum_2004} and of the $\sigma$ fields~\cite{ginsparg_applied_1988,di_francesco_conformal_2005}, we hence obtain
\begin{equation}
\label{eq:correlation_functions}
\begin{aligned}
    G_{r}(\tau) &=&
    \langle e^{-\frac{i}{2}\phi_{r}(\tau)} e^{\frac{i}{2}\phi_{r}(0)} \rangle_0 
    \langle \sigma_{r} (\tau) \sigma_{r} (0) \rangle_0 \\
    &=& \left[\frac{\pi T_{r} \tau_c}{i \sinh \left( \pi T_{r} \left( \tau - i \tau _c \right) \right)} \right]^{2 \delta}
    \end{aligned}
\end{equation}
where $\tau_c$ is a short time cutoff, and $\delta = \delta_\phi + \delta_\sigma$ is the scaling dimension, given by the sum of the scaling dimensions for the boson field and the $\sigma$ field. For systems without effective interactions between the quasiparticles, $\delta_\phi = 1/16$, and $\delta_\sigma = 1/16*{N_\chi}$ \cite{yang_influence_2013}. From this point in the calculation, the only signifier that we are considering the $\nu = 5/2$ state and not any other FQH state are the explicit values of $e^*=e/4$ and $s_0 = \log (\sqrt{2})$.

We emphasize that $I_{\rm J}$ is consistent with previous works which have considered heat current in the absence of an internal entropy term \cite{ronetti_hong-ou-mandel_2019,vannucci_minimal_2017,ebisu_fluctuations_2022} and not in the shot noise limit. (Results for different limits are discussed in App.~\ref{app:limits}.) These expressions are valid for all types of processes ($p$, $h$, $hp$, $ph$ etc.). Only the internal entropy term $I_{\rm S}$ is different between them.

As noted above, it is useful to consider the shot noise limit of $V \gg T_R,T_L$, as this suppresses some of the transport processes. Formally, this is incorporated into Eq.~\eqref{eq:generalcurrents} by expanding $G_{R/L} \approx \left[\tau_c/(i\tau +\tau_c)\right]^{2\delta} + \mathcal{O}(T_{R/L}^2)$. One immediately finds that $I_{\rm J}=\frac{1}{2} I V$ through integration by parts. This term corresponds to Joule heating, with a pre-factor of $1/2$ as expected due to the presence of a full sea. As can be seen, by the expansion of the correlation functions, the sub-leading terms are to quadratic order in both temperatures and thus are well distinguished from the $T$-linear term $I_S$.

To obtain the term $I_{\rm S}=\frac{I}{e^*} T_L s_0$, we have to assume that only one type of charge arrives at the left movers' edge. As noted, the conformal low-energy theory used is limited, as we have no distinction between creating a quasihole and annihilating a quasiparticle. Collecting the expression for $I_S$ and $I_J$ we arrive at our main result in Eq.~\eqref{eq:IQ}.

\textit{Discussion.}---In this work, we search for an available experimental probe to identify the entropic signatures of non-Abelian quasiparticles in fractional quantum Hall states, focusing on the state with filling fraction $\nu=5/2$. We propose to measure the heat current, which has been the focus of much experimental \cite{banerjee_observed_2017,banerjee_observation_2018,melcer_absent_2022,melcer_direct_2023,melcer2023heat,dutta_distinguishing_2022,dutta_isolated_2022,srivastav2019,Srivastav2021,Kumar2021b} and theoretical \cite{cappelli_thermal_2002, viola_thermoelectric_2012, Park_2020, hein_2023, wei_thermal_2023} attention in regards to this state. Our proposal focuses on the heat current that tunnels through a weak constriction between FQH edges. We find that measuring this heat current and the resulting charge current reveals the contribution of the internal entropy of the non-Abelian quasiparticles.

Surprisingly, we find that the standard low-energy effective field theory \cite{wen_quantum_2004} used to describe FQH edge modes and relate the quantum dimension of the primary fields to their internal entropy is insufficient for calculating the entropy contribution to the heat current. This is due to the presence of different contributions of electrically similar yet entropically distinguishable tunneling processes, which the low-energy conformal field theory inherently treats as identical. 

We therefore identify the different processes, and sum their contribution to the entropy transfer in conjunction with the typical low-energy theory. We thus find the signature exists in the ratio between the tunneling heat and charge currents. To obtain this signature, we find the system should be tuned to the shot noise limit, $V \gg T_R,T_L$, and the tunneling has to be dominated by a single type of quasiparticle. 

It appears that the specific conditions under which only one type of quasiparticle dominates tunneling depend on many details that are specific to the exact realization of the quantum Hall state. However, we speculate that in the shot noise limit, when one of the edges is connected to a reservoir at a voltage greater than the temperature, and the bulk and the other edge are grounded, there might be a transfer of charge and energy by one type of quasiparticles, and the anomalous internal entropy can be determined.

\textit{Acknowledgements.}--- We are grateful for insightful discussions with Matteo Acciai, Eduardo Fradkin, Moty Heiblum, Ronni Himmel, June-Young Lee, Bernd Rosenow, Hueng-Sun Sim  and Christian Sp\aa{}nsl\"att.
This work received support from the NSF-BSF award DMR-2310312, the European Union’s Horizon 2020 research and innovation programme (grant agreements LEGOTOP No. 788715), the DFG CRC SFB/TRR183, the Simons Foundation, the NSF DMR Grant number 1839271, and the IQIM, an NSF Physics Frontiers Center, as well as AFOSR MURI FA9550-22-1-0339, and the KAKENHI-PROJECT-23H01097. N.S. was supported by the Clore Scholars Programme. Part of this work was done at the Aspen Center for Physics, which is supported by the NSF grant PHY-1607611.

\bibliography{main}

\appendix
\section{Technical derivations}

In this appendix, we show how internal entropy enters into the heat current operator on thermodynamic grounds. We then proceed to derive Eq.~\eqref{eq:generalcurrents} of the main text, and subsequently, by taking the appropriate shot noise limit (with temperature much smaller than the potential), Eq.~\eqref{eq:IQ}. Finally, we discuss a few other limits of interest that do not result in Eq.~\eqref{eq:IQ}.

\subsection{Thermodynamic prelude}
\label{app:Thermo}
Here, we derive the heat current on thermodynamic grounds, following the example of, e.g.,  Refs.~\cite{Schroeder_thermal,hoch_statistical_2021}. The heat in a reservoir is given by $Q \equiv T S$. Thus, to determine the heat current, assuming temperatures are kept constant, one must find the entropy current. Using the chain rule, we obtain
\begin{equation}
    \label{eq:entropy_chain}
    \frac{d S}{d t}= \left(\frac{\partial S}{\partial U}\right)_{N,T,\Omega}\frac{\partial U}{\partial t} + \sum_i \left(\frac{\partial S}{\partial N_i}\right)_{U,T,\Omega,N_{j\neq i}}\frac{\partial N_i}{\partial t},
\end{equation}
where $\Omega$ is the volume of the reservoir, $U$ is energy, and $N_i$ is particle number for the $i$th particle type. Since we assume tunneling is rare, and thus each reservoir approximately maintains equilibrum, we can identify $\left(\frac{\partial S}{\partial U}\right)_{N,T,\Omega} \equiv \frac{1}{T}$ and $\left(\frac{\partial S}{\partial N}\right)_{U,T,\Omega,N_{j\neq i}} \equiv -\frac{\mu_i}{T}$, giving the standard expression
\begin{equation}
    \label{eq:heat_current}
    \frac{d Q}{d t}= \frac{\partial U}{\partial t} - \sum_i \mu_i \frac{\partial N_i}{\partial t}.
\end{equation}

The heat current thus decomposes to two separate calculations, of energy current and particle current. The particle current is then multiplied by the chemical potential: crucially, this is the chemical potential of the reservoir in which the current is calculated. There is subtlety, however, in the definition of these chemical potentials, as it includes multiple components. For our purposes, the two components of interest are the electro-chemical component and the internal entropy component. 

We now demonstrate this using an explicit derivation of the chemical potential of a dilute gas, where we can work out explicitly the role that the internal entropy plays. Let us assume this ideal gas consists of $N$ particles in a volume $\Omega$, and internal entropy $s_0$.  The partition function of the reservoir is given by $\mathcal{Z} = z^N/N!,$ where $z$ is  the single-particle partition function. This is given by
\begin{equation}
    \label{eq:single_partition}
    z = \int_\Omega \frac{d^3 \mathbf{x} d^3 \mathbf{p}}{(2\pi \hbar)^3} e^{-\beta \left(\frac{\mathbf{p}^2}{2m}-e^* V\right)} \times z_{\textrm{int}}.
\end{equation}
Here, $\beta$ is inverse temperature, $m$ is particle mass, $e^*$ is particle charge, and $V$ is voltage. An important emphasis must be placed on the term $z_{\textrm{int}}$, which is the partition function of the particle's internal degrees of freedom \cite{Schroeder_thermal,hoch_statistical_2021}. These may include rotational and vibrational degrees of freedom, or orbital energy levels. For our purposes, in which we consider particles with an internal degeneracy of $d$, we simply replace $z_{\textrm{int}}=d\equiv e^{s_0}$. We emphasize that this is a classical degeneracy; thus, one can not use this treatment to discuss spinful fermions by reappropriating this derivation with $d$ denoting spin degeneracy. 

Now completing the Gaussian integral, and using the Stirling approximation, we find the free energy $F = -T \log \mathcal{Z}$ is given by
\begin{equation}
    \label{eq:free_energy}
    \mathcal{F} = -T N \log \frac{\Omega}{N} \sqrt{\left( \frac{2\pi m T}{\hbar^2}\right)^3} -TN +N e^* V-TN s_0. 
\end{equation}
The chemical potential is then given by $\left(\partial \mathcal{F}/\partial N \right)$, and is
\begin{equation}
    \label{eq:chemical_potential}
    \mu = -T \log \frac{\Omega}{N} \sqrt{\left( \frac{2\pi m T}{\hbar^2}\right)^3} +e^* V -T s_0. 
\end{equation}

The above expression naturally decomposes to three main components of the chemical potential: a kinetic term, $\mu_k \equiv -T \log \frac{\Omega}{N} \sqrt{\left( \frac{2\pi m T}{\hbar^2}\right)^3}$; an electrochemical term, $\mu^e \equiv e^* V$; and an internal entropy term, $\mu^s \equiv - T s_0$. We focus on the latter two parts, as the kinetic term will vanish as the temperature is taken deep into the quantum degenerate regime, and by the third law of thermodynamics will vanish faster than $T$. It is immediately apparent that the issue of internal entropy must be carefully accounted for when inserting the chemical potentials of Eq.~\eqref{eq:chemical_potential} in the explicit heat current calculation of Eq.~\eqref{eq:heat_current}.

In our problem, the reservoirs in question are the equilibrated contacts along the relevant quantum Hall edges. The presence of the internal entropy term $\mu^s$ in the chemical potential manifests in an entropy dependent term in the heat current entering the cold reservoir of the left-moving edge, $I_Q \equiv \frac{d Q_L}{dT}$. 

In order to isolate this term from non-universal prefactors such as the tunneling probability through the bridge, we will also wish to consider the electric current.
\begin{equation}
    \label{eq:electric_current}
    I \equiv \sum_i e^*_i\frac{\partial N_{L,i}}{\partial t}.    
\end{equation}
It will be convenient for calculation purposes to re-define Eq.~\eqref{eq:heat_current} in a similar form. We decompose the heat current into two contributions, $I_Q \equiv I_J + I_S$, where $I_J$ is the heat current which is driven by the energetics of the system, and $I_S$ is the heat current which is driven by the contributions of internal entropy. These can be written as 
\begin{subequations}
    \label{eq:heat_en_int}
    \begin{align}
            I_J & \equiv \sum_i \left( \varepsilon_i-\mu_i^e \right)\frac{\partial N_{L,i}}{\partial t}, \\
            I_S & \equiv - \sum_i \mu_i^s\frac{\partial N_{L,i}}{\partial t},
    \end{align}
\end{subequations}    
where $\varepsilon_i \equiv (\frac{\partial U}{\partial t})_i / (\frac{\partial N_i}{\partial t})$ is defined as the average energy carried by each tunneling quasiparticle of type $i$. We note that in the absence of an internal entropy term, the heat current is given entirely by $I_J$. This has been the term calculated in several works considering heat current in FQH states \cite{vannucci_minimal_2017,ronetti_hong-ou-mandel_2019,ebisu_fluctuations_2022}.

\subsection{Quantum Derivation}
\label{app:quantum}
The currents $I_{\rm J}$ and $I_{\rm S}$ quantities can now be calculated on a quantum level for the problem at hand using a Hamiltonian formalism. The Hamiltonian of a general system of two quantum Hall edges connected by a tunneling term can generally be written as
\begin{equation}
    \label{eq:Hamiltonian}
    \mathcal{H}= H_R + H_L + H_T.
\end{equation}
Here, $H_{R/L}$ is the Hamiltonian of the right-/left-moving edge, such as the Hamiltonian given in Eq.~\eqref{eq:Hamiltonian_five_halves} of the main text for the $\nu = 5/2$ FQH state. $H_T$ describes the tunneling. For a single type of tunneling quasiparticle/quasihole, we define this as
\begin{equation}
    \label{eq:tunneling_hamiltonian2}
    H_T = \Gamma \Psi^\dagger_R(x=0) \Psi_L(x=0) + \textrm{h.c.},
\end{equation}
where $\Gamma$ is a tunneling amplitude, assumed to be real, and $\Psi_{R/L}$ annihilates a quasipartilce at the right-/left- moving edge. We remark that if we simultaneously identify $\Psi_{R/L}$ as a quasihole creation operator, then a Hamiltonian of this nature may describe the several different processes described in the main text; for example, both right-to-left tunneling of a quasiparticle and left-to-right tunneling of a quasihole. As noted in the main text, these processes are identical for charge carrying considerations, but carry different entropy.

We now idenfity the total energy along the left-moving edge as $U_L = H_L$, and the total particle number along the left-moving edge as $N_L = \int dx \Psi^\dagger_L(x) \Psi_L(x)$. The particle-tunneling and energy-tunneling operators are now defined directly via commutation with the tunneling Hamiltonian,
\begin{align}
    \label{eq:heat_current_operators}
        \hat{\dot{ N}}_L & \equiv i \left[ H_T, \hat{N}_L \right] = i \Gamma \left \{ \Psi^\dagger_R \Psi_L - \Psi^\dagger_L \Psi_R \right \}, \\
        \hat{\dot{U}}_L & \equiv i \left[ H_T, \hat{U}_L \right] = - \Gamma \left \{ \Psi^\dagger_R \left(\partial_t \Psi_L\right) + \left(\partial_t\Psi^\dagger_L\right) \Psi_R \right \}, \nonumber
\end{align}
where we have suppressed the position coordinates for brevity. 

The charge current and the energy current which tunnel at the QPC are now given by the expectation values of these operators, i.e., $\frac{\partial N_L}{\partial t} \equiv \langle \hat{\dot{ N}}_L \rangle $ and $\frac{\partial U_L}{\partial t} \equiv \langle \hat{\dot{U}}_L \rangle $. These are calculated perturbatively in the tunneling term $H_T$. To leading order, these expectation values factorize for the right- and left-moving edges, with each sector calculated with respect to its unperturbed Hamiltonian, $H_{R/L}$. We perform this calculation to leading order in the tunneling amplitude, $\mathcal{O} \left(\Gamma^2\right)$ by using the Kubo or Keldysh formalism, as they are equivalent to this order~\cite{martin_noise_2005}. 

Defining $G_{R/L}^{>}(t-t^\prime) \equiv \langle \Psi^\dagger_{R/L}(t) \Psi_{R/L}(t^\prime)\rangle_0$ and $G_{R/L}^{<}(t-t^\prime) \equiv \langle \Psi_{R/L}(t) \Psi^\dagger_{R/L}(t^\prime)\rangle_0$, where $\langle \cdots \rangle_0$ denotes taking an expectation value with respect to the unperturbed Hamiltonian $H_{R/L}$, one obtains after standard manipulations \cite{martin_noise_2005,schiller_extracting_2022,ebisu_fluctuations_2022}
\begin{align}
    \label{eq:kubos}
        \frac{\partial N_L}{\partial t} & = 
        \Gamma^2 \hspace{-0.4em} \int_{-\infty}^{\infty} \hspace{-0.5em} d\tau 
        \! \left\{ G_{R}^{>} \!\left( \tau \right) G_{L}^{<} \!\left( \tau \right) \! 
        - \! G_{R}^{<} \!\left( \tau \right) G_{L}^{>} \!\left( \tau \right) \right \} \!, \\
        \frac{\partial U_L}{\partial t} &= 
        i \Gamma^2 \hspace{-0.4em}\int_{-\infty}^{\infty} \hspace{-0.5em} d\tau 
        \! \left\{ G_{R}^{>} \!\left( \tau \right) \partial_\tau G_{L}^{<} \!\left( \tau \right) \! + \! G_{R}^{<} \!\left( \tau \right) \partial_\tau G_{L}^{>} \!\left( \tau \right) \right \} \! . \nonumber
\end{align}

To proceed in this calculation, we require proper knowledge of the Green's functions. We find these using the operators of Eq.~\eqref{eq:excitations} in the main text. The two separate fields factorize, giving
\begin{align*}
    G_{R/L}^{>}(\tau) & = 
    \langle e^{-\frac{i}{2}\phi_{R/L}(\tau)} e^{\frac{i}{2}\phi_{R/L}(0)} \rangle_0 
    \langle \sigma_{R/L} (\tau) \sigma_{R/L} (0) \rangle_0, \\ 
    G_{R/L}^{<}(\tau) & = 
    \langle e^{\frac{i}{2}\phi_{R/L}(\tau)} e^{-\frac{i}{2}\phi_{R/L}(0)} \rangle_0  
    \langle \sigma_{R/L} (\tau) \sigma_{R/L} (0) \rangle_0. 
\end{align*}
Using known values for correlations of the boson fields \cite{giamarchi_quantum_2003,wen_quantum_2004} and of the $\sigma$ fields \cite{ginsparg_applied_1988,di_francesco_conformal_2005}, we hence obtain
\begin{equation}
    \begin{aligned}
    \label{eq:correlation_functions2}
    G^{\gtrless}_{R/L}(\tau) & = e^{i \pm e^* V_{R/L} (t-t^\prime)} G_{R/L}(\tau), \\
    G_{R/L}(\tau) & = \left[\frac{\pi T_{R/L} \tau_c}{i \sinh \left( \pi T_{R/L} \left( \tau - i \tau _c \right) \right)} \right]^{2 \delta},
    \end{aligned}
\end{equation}
where $\tau_c$ is a short time cutoff, and $\delta = \delta_\phi + \delta_\sigma$ is the scaling dimension, given by the sum of the scaling dimensions for the boson field and the $\sigma$ field. For systems without effective interactions between the quasiparticles, $\delta_\phi = 1/16$, and $\delta_\sigma = 1/16*{N_\psi}$ \cite{yang_influence_2013}. From this point on in the calculation, the only signifier that we are considering the $\nu = 5/2$ state and not any other FQH state are the explicit values of $e^*=e/4$ and $s_0 = \log (\sqrt{2})$.

We emphasize a few important notes. First, is that the only difference between the greater and lesser Green's functions are the sign of the electro-chemical potentials. This is because voltage is the only mechanism our current system has which breaks particle-hole symmetry. Second, the oscillatory term is dictated by the electro-chemical potential, and not the entire chemical potential. This can be derived through arguments of gauge invariance~\cite{chamon_tunneling_1995}.

We now wish to use these correlation functions to obtain a closed expression for the heat and charge current. Plugging Eq.~\eqref{eq:correlation_functions} into Eqs.~\eqref{eq:electric_current} and \eqref{eq:heat_en_int}, we find
\begin{subequations}
    \label{eq:generalcurrents2}
    \begin{align}
        I & = 2 i e^* \Gamma^2 \int_{-\infty}^\infty d\tau \sin \left(e^* V \tau \right) G_R(\tau) G_L(\tau), \\
        I_J & = 2 i \Gamma^2 \int_{-\infty}^\infty d\tau \cos \left(e^* V \tau \right) G_R(\tau) \partial_\tau G_L(\tau).
    \end{align}
\end{subequations}
We emphasize that $I_J$ is consistent with previous works which have considered heat current in the absence of an internal entropy term \cite{ronetti_hong-ou-mandel_2019,vannucci_minimal_2017,ebisu_fluctuations_2022}.

Crucially, for a single tunneling quasiparticle, the form of Eq.~\eqref{eq:heat_en_int} reduces to~$I_S = -\mu^s \dot{N} = -\frac{\mu^s}{e^*}I$. As we argue in the main text, we anticipate a single quasiparticle to dominate tunneling in the shot noise limit. An added benefit of this limit is the simple reduction of the energy-driven term into $I_J= IV/2$; this can be easily shown by expanding $G_{R/L} \approx \left[\tau_c/(i\tau +\tau_c)\right]^{2\delta} + \mathcal{O}(T_{R/L}^2)$, and proceeding with integration by parts of Eq.~\eqref{eq:generalcurrents2}. This, combined with the identification of $\mu^s = - T_L s_0$ for quasiparticles (or $\mu^s = T_L s_0$ for quasiholes, if tunneling is dominated by them) allows us to complete the derivation of our main equation.

\subsection{Alternate limits}
\label{app:limits}

The first limit which must receive treatment is that in which the Hamiltonian is not dominated by a single tunneling process, but by some combination of the four processes discussed in the main text. In this case, one may proceed with a similar perturbative treatment by multiplying the tunneling Hamiltonian for each process,
\begin{align}
    H_T &\mapsto \sum_{i} H_T^{i}; \qquad i \in \left(p, h, hp, ph \right) \\
    H_T^{i} &= \Gamma_{(i)} \psi^\dagger_{R,(i)}\psi_{L,(i)}+ \textrm{h.c.}.
\end{align}

The currents are then derived by repeating the steps of the previous section while carefully accrediting each process with the appropriate entropic contribution. One then obtains
\begin{subequations}
    \label{eq:fourprocesscurrents}
    \begin{align*}
        I & = 2 i e^* \left(\sum_{i} \Gamma_{(i)}^2\right) \int_{-\infty}^\infty d\tau \sin \left(e^* V \tau \right) G_R(\tau) G_L(\tau), \\
        I_J & = 2 i \left(\sum_{i} \Gamma_{(i)}^2\right) \int_{-\infty}^\infty d\tau \cos \left(e^* V \tau \right) G_R(\tau) \partial_\tau G_L(\tau). \\
        I_S & = 
        \frac{\Gamma_p^2 - \Gamma_h^2 + \Gamma_{hp}^2 - \Gamma_{ph}^2}
        {\Gamma_p^2 + \Gamma_h^2 + \Gamma_{hp}^2 + \Gamma_{ph}^2} \;
        \frac{T_L s_0}{e^*} I.
    \end{align*}
\end{subequations}
While we re-obtain the previous limit for a single dominant process, in the fully particle-hole symmetric case, in which $\Gamma_p = \Gamma_h$ and $\Gamma_{hp}=\Gamma_{ph}$, the entropic contribution disappears. It is thus crucial to perform the measurement in the shot noise limit, where particle-hole symmetry is most violently broken.

One can also consider the alternative regime of temperature-only biasing, $V = 0$, which trivially gives a charge current of $I=0$.
As has been shown in Ref.~\cite{ebisu_fluctuations_2022}, an explicit calculation gives $I_J = A(T,\delta) \Delta T$, where the coefficient $A(T,\delta)$ depends on the temperature and the scaling dimension. However, as no net particle number of any type tunnels between the edges, we will not obtain a term which depends on the internal entropy.

\end{document}